\documentclass[12pt]{iopart}

\usepackage{iopams} 

\usepackage{graphicx} 
\begin{document}

%\title[Measuring the equation of state of a homogeneous ultracold gas]{Measuring the equation of state of a homogeneous ultracold gas}
\title[The equation of state of ultracold Bose and Fermi gases: a few examples]{The equation of state of ultracold Bose and Fermi gases: a few examples}

\author{Sylvain Nascimb\`ene, Nir Navon, Fr\'ed\'eric Chevy, Christophe Salomon}

\address{Laboratoire Kastler Brossel, CNRS, UPMC, \'Ecole
Normale Sup\'erieure, \\
24 rue Lhomond, 75231 Paris, France}
\ead{sylvain.nascimbene@lkb.ens.fr}
\begin{abstract}
%We describe a new method for determining the equation of state of a homogeneous ultracold gas from \emph{in situ} images of trapped gases. It is based on a measurement of the local pressure inside the gas, In the local density framework, 
We describe a powerful method for determining the equation of state of an ultracold gas from \emph{in situ} images. The method provides a measurement of the local pressure of an harmonically trapped gas and we give several applications to Bose and Fermi gases. We obtain the  grand-canonical equation of state of a spin-balanced Fermi gas with resonant interactions as a function of temperature \cite{nascimbene2009eos}.
%We describe the procedure used to obtain the equation of state of a spin-balanced Fermi gas with resonant interactions. 
We compare our equation of state with an equation of state measured by the Tokyo group in \cite{horikoshi2010measurement}, that reveals a significant difference in the high-temperature regime. The normal phase, at low temperature, is well described by a Landau Fermi liquid model, and we observe a clear thermodynamic signature of the superfluid transition.
%At low temperature the equation of state is well described,  in the normal phase, as a Landau Fermi liquid, and we observe a thermodynamic signature of the superfluid transition.
In a second part we apply the same procedure to Bose gases.
From a single image of a quasi ideal Bose gas we determine the equation of state from the classical to the condensed regime. Finally the method is applied to a Bose gas in a 3D optical lattice in the Mott insulator regime. Our equation of state  directly reveals the Mott insulator behavior and is suited to investigate finite-temperature effects. 
\end{abstract}

\maketitle

%\tableofcontents
\section{Introduction}

Ultracold gases are a privileged tool for the simulation in the laboratory of model Hamiltonians relevant in the fields of condensed matter, astrophysics or nuclear physics \cite{bloch2008many}. As an example, thanks to the short-range character of interactions, ultracold Fermi mixtures prepared around a Feshbach resonance  mimic the behavior of neutron matter in the outer crust of neutron stars \cite{bertsch2001many,gezerlis2008strongly}. For cold atoms, the density inhomogeneity induced by the trapping potential has long made the connection between the Hamiltonian of a homogeneous system and an ultracold gas indirect. Early experimental thermodynamic studies have provided global quantities averaged over the whole trapped gas, such as total energy and entropy \cite{stewart2006potential,luo2007measurement}, collective mode frequencies \cite{altmeyer2007precision}, or
 radii of the different phases that may be observed in an imbalanced Fermi gas \cite{partridge2006pap,zwierlein2006direct,nascimbene2009pol}. 
Reconstructing the equation of state of the homogeneous gas then requires to deconvolve the effect of the trapping potential, a delicate procedure that has not been done so far. However, the gas can often be considered as locally homogeneous (local density approximation (LDA)), and careful analysis of \emph{in situ} density profiles can directly provide the equation of state of a homogeneous gas \cite{shin2008des,ho2009opdtq,nascimbene2009eos,navon2010eos}.
In the case of two-dimensional gases, \emph{in situ} images taken along the direction of tight confinement  obviously give access to the surface density \cite{hadzibabic2006berezinskii,cladé2009observation,gemelke2009situ,bakr2009quantum} and thus to the equation of state \cite{rath2010equilibrium}. For three-dimensional gases, imaging leads to an unavoidable integration along the line of sight. As a consequence, inferring local quantities is not straightforward. Local density profiles can be computed from a cloud image using an inverse Abel transform for radially symmetric traps \cite{shin2008pd}. A more powerful method was suggested in \cite{ho2009opdtq} and implemented in \cite{nascimbene2009eos,navon2010eos}: as explained below, for a harmonically trapped gas the local pressure is simply proportional to the integrated \emph{in situ} absorption profile. 
Using this method, the low-temperature superfluid equation of state for balanced and imbalanced Fermi gases have been studied as a function of interaction strength \cite{nascimbene2009eos,navon2010eos}. 
In this paper we describe in more detail the procedure used to determine the equation of state of a spin-unpolarized Fermi gas in the unitary limit \cite{nascimbene2009eos}. We compare our data with recent results from the Tokyo group \cite{horikoshi2010measurement}, and reveal a significant discrepancy in the high-temperature regime. In a second part we apply the method to ultracold Bose gases. From an \emph{in situ} image of $^7$Li, we obtain the equation of state of a weakly-interacting Bose gas. Finally, analyzing experimental profiles of a Bose gas in a deep optical lattice \cite{fölling2006formation}, we observe clear thermodynamic signatures of the Mott insulator phases.

\section{Measurement of the local pressure inside a trapped gas}
In the grand-canonical ensemble, all thermodynamic quantities of a macroscopic system can be derived from the equation of state $P=f(\mu,T)$ relating the pressure $P$ to the chemical potential $\mu$ and temperature $T$.
$P$ can be straigthforwardly deduced from integrated \emph{in situ} images.

\begin{figure}[t]
      \begin{center}
   \includegraphics[width=1\linewidth]{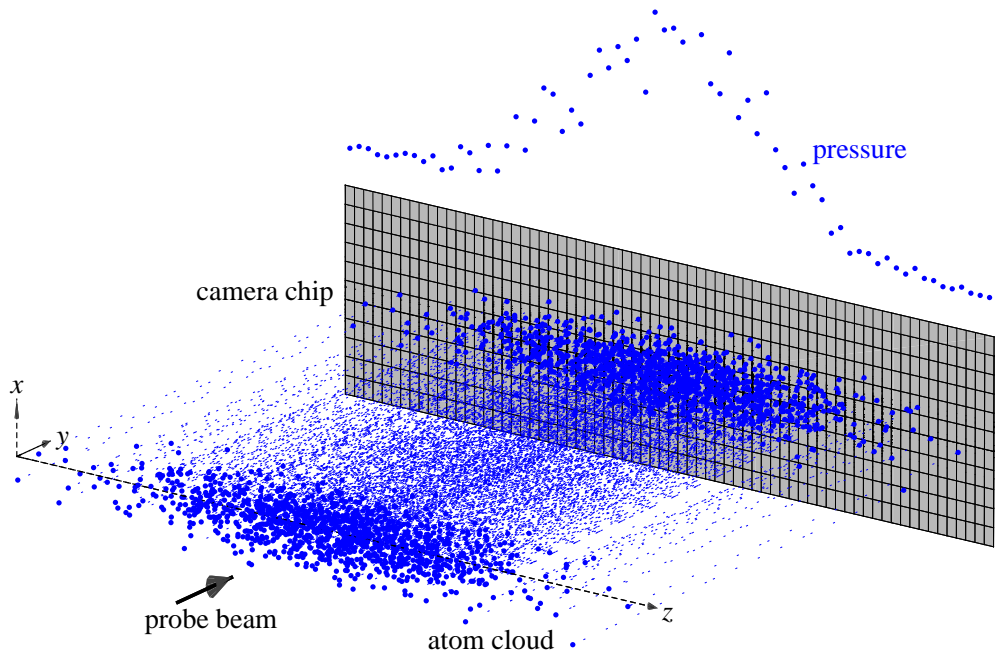}
   \vspace{-1cm}
      \end{center}
   \caption{Scheme of the local pressure measurement: the absorption of a probe beam propagating along the $y$ direction provides a 2D image on the CCD camera. Integration of this image along $x$ provides the doubly-integrated density profile $\overline{n}(z)$ and, using equation (\ref{Pressure}), the pressure profile along the $z$ axis.
\label{Schema_Image}}
\end{figure}

Consider first a single-species ultracold gas, held in a cylindrically symmetric harmonic trap whose frequencies are labeled $\omega_x=\omega_y\equiv\omega_r$ in the transverse direction, and $\omega_z$ in the axial direction. 
Provided that the local density approximation is satisfied, the gas pressure along the $z$ axis is given by \cite{ho2009opdtq}:
\begin{equation}\label{Pressure}
P(\mu_z,T)=\frac{m\omega_r^2}{2\pi}\overline{n}(z),
\end{equation}
where $\overline{n}(z)=\int\mathrm{d}x\,\mathrm{d}y\,n(x,y,z)$ is the doubly-integrated density profile, $\mu_z=\mu^0-\frac{1}{2}m\omega_z^2z^2$ is the local chemical potential on the $z$ axis, $\mu^0$ is the global chemical potential. $\overline{n}(z)$ is obtained from an \emph{in situ} image taken along $y$, by integrating the optical density along $x$ (see Fig.\ref{Schema_Image}).  As described below, if one independently determines the temperature $T$ and chemical potential $\mu^0$, then
each pixel row of the absorption image at a given position $z$, provides an experimental data point for the grand-canonical equation of state $P(\mu_z,T)$ of the \emph{homogeneous} gas. The large amount of data obtained from several images allows one to perform an efficient averaging, leading to a low-noise equation of state.

This formula is also valid in the case of a two-component Fermi gas with equal spin populations if $\overline{n}(z)$ is the total integrated density.
The method can be generalized to multicomponent Bose and Fermi gases, as first demonstrated on spin-imbalanced Fermi gases in \cite{nascimbene2009eos,navon2010eos}. 

\section{Thermodynamics of a Fermi gas with resonant interactions}
In this section we describe the procedure used in \cite{nascimbene2009eos} to determine the grand-canonical equation of state of a homogeneous and unpolarized Fermi gas with resonant interactions ($a=\infty$). We also compare our data with recent measurements from the Tokyo group \cite{nascimbene2009eos,horikoshi2010measurement}. We then study the physical content of the equation of state at low temperature. 

\subsection{Grand-canonical equation of state}
\begin{figure}[ht!]
      \begin{center}
   \includegraphics[width=1\linewidth]{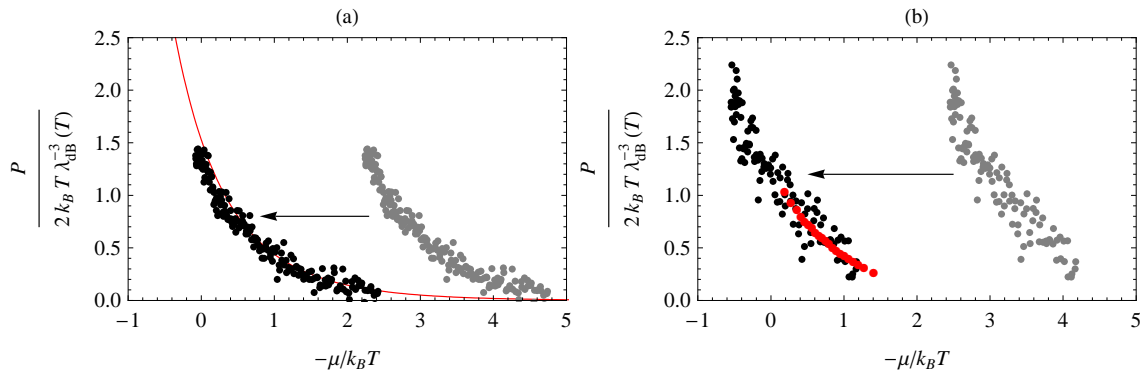}
   \vspace{-1cm}
      \end{center}
   \caption{Determination of $\mu^0$: we plot the data from an \emph{in situ} image as $P/2k_BT\lambda_{dB}^{-3}$ versus $-\mu/k_BT=V(z)/k_BT-\mu^0/k_BT$ (black points). A wrong choice of $\mu^0$ corresponds in this representation to a translation of the data in abscissa. We adjust $\mu^0$ so that the wings of the pressure profile match a reference equation of state (in red). (a) For high-temperature clouds, we use the second-order virial expansion (\ref{virial2}) (b) For a lower temperature pressure profile, we minimize its distance with the averaged equation of state deduced from higher temperature images (in red)  in the overlap region.
\label{construction}}
\end{figure}

In the grand-canonical ensemble, the equation of state of a spin-unpolarized Fermi gas in the unitary limit, can be written as
\begin{equation}\label{hT}
P(\mu,T)=P^{(0)}(\mu,T)h_T(\zeta),
\end{equation}
where $P^{(0)}(\mu,T)$ is the pressure of a non-interacting two-component Fermi gas and $\zeta=\exp(\-\mu/k_BT)$ is the inverse fugacity. Since $P^{(0)}(\mu,T)$ is known, the function $h_T(\zeta)$ completely determines the equation of state $P(\mu,T)$. Let us now describe the procedure used to measure it. The pressure profile of the trapped gas along the $z$ axis is directly obtained from its \emph{in situ} image using equation (\ref{Pressure}). One still has to know the value of the temperature $T$ and global chemical potential $\mu^0$ in order to infer $h_T(\zeta)$. We use a small amount of $^7$Li atoms, at thermal equilibrium with the $^6$Li component, as a thermometer. We then extract $\mu^0$ from the pressure profile, by comparison in the cloud's wings with a reference equation of state. For high-temperature clouds ($k_BT>\mu^0$), we choose $\mu^0$ so that the wings of the pressure profile match the second-order virial expansion \cite{ho2004high} (see Fig.\ref{construction}a):
\begin{equation}\label{virial2}
P(\mu,T)=\frac{2k_BT}{\lambda_{dB}^3(T)}\left(e^{\mu/k_BT}+\frac{4}{3\sqrt{2}}e^{2\mu/k_BT}+\ldots\right).
\end{equation}
For colder clouds, the signal-to-noise ratio is not good enough, in the region where (\ref{virial2}) is valid, to extract $\mu^0$ using the same procedure. We thus rather use the equation of state determined from all images previously treated as a reference, since it is accurate on a wider parameter range than (\ref{virial2}) (see Fig.\ref{construction}b). We then iterate this procedure at lower and lower temperatures, eventually below the superfluid transition. By gathering the data from all images and statistical averaging, we obtain a low-noise equation of state in the range $0.02<\zeta<5$ (see Fig.\ref{Comparison_Mukaiyama}a). 

%In \cite{nascimbene2009eos} we measured a grand-canonical equation of state, $P(\mu,T)/P^{(0)}(\mu,T)$ as a function of $\zeta=e^{-\mu/k_BT}$, from pressure profiles extracted from \emph{in situ} images using equation (\ref{Pressure}) (see Fig.\ref{Comparison_Mukaiyama}a). Here $P^{(0)}(\mu,T)$ is the pressure of a non-interacting two-component Fermi gas. The cloud temperature $T$ is measured on a weakly-interacting $^7$Li gas at thermal equilibrium with the $^6$Li component. The global chemical potential $\mu^0$ is determined from the wings of the pressure profile, by comparison with the high-temperature virial expansion up to second-order \cite{ho2004high}:
%\[
%P(\mu,T)=\frac{2k_BT}{\lambda_{dB}^3(T)}\left(e^{\mu/k_BT}+\frac{4}{3\sqrt{2}}e^{2\mu/k_BT}\right).
%\]

\subsection{Canonical equation of state}
\begin{figure}[ht!]
      \begin{center}
   \includegraphics[width=1\linewidth]{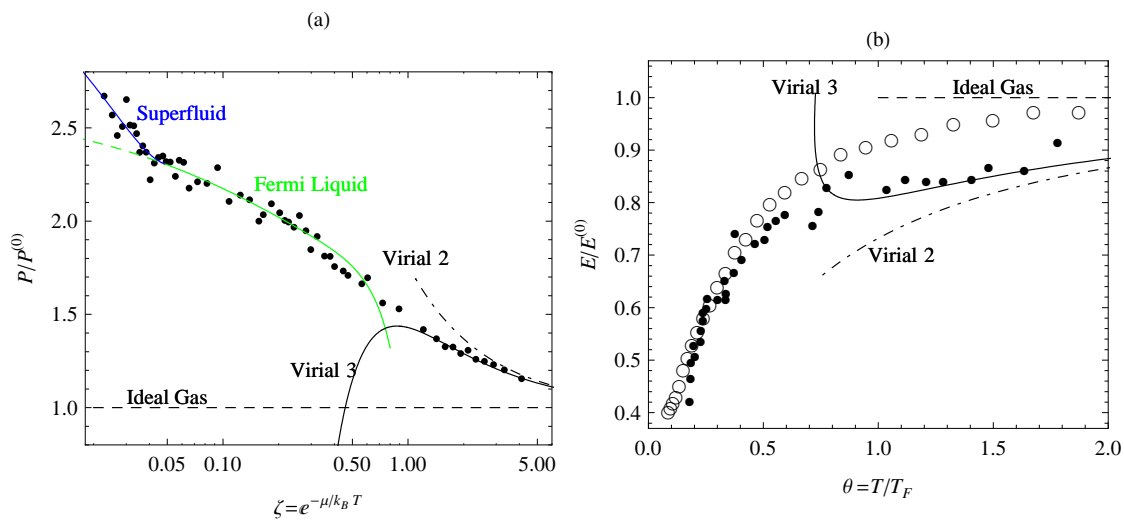}
   \vspace{-1cm}
      \end{center}
   \caption{(a) Grand-canonical equation of state of a two-component Fermi gas with resonant interactions from \cite{nascimbene2009eos} (black dots). (b) Canonical equation of state from the Tokyo group \cite{horikoshi2010measurement} (open circles) and from the ENS group (black dots). The dashed black line is the ideal gas equation of state, the dot-dashed (solid) black line is the second- (third-) order virial expansion, the solid green line is the Fermi liquid equation (\ref{eq_Fermi}) and the solid blue line is the fit function (\ref{eq_c}) in the superfluid phase. The superfluid transition occurs at $\zeta=0.05$.
\label{Comparison_Mukaiyama}}
\end{figure}

In \cite{horikoshi2010measurement} a canonical equation of state $E(n,T)$ expressing the energy $E$ as a function of density and temperature was measured using fits of absorption images taken after a short time-of-flight. \emph{In situ} density profiles were deduced by assuming a hydrodynamic expansion. The temperature was extracted from the cloud's total potential energy at unitarity, using the experimental calibration made in \cite{luo2007measurement}. In Fig.\ref{Comparison_Mukaiyama}b the data from \cite{horikoshi2010measurement} is plotted as $E(n,T)/E^{(0)}(n,T)$ as a function of $\theta=T/T_F$, where $n$ is the total atom density, $T_F$ is the Fermi temperature, and $E^{(0)}(n,T)$ is the energy of a non-interacting Fermi mixture.

The comparison between the two equations of state requires to express our data in the canonical ensemble. The density $n=\partial P/\partial\mu|_T$ is calculated by taking a discrete derivative, and we obtain the black points in Fig.\ref{Comparison_Mukaiyama}b. While the two sets of data are in  satisfactory agreement in the low-temperature regime $T/T_F<0.4$, they clearly differ in the high-temperature regime. The disagreement of the data from \cite{horikoshi2010measurement} with the second- and third-order virial expansions calculated in \cite{ho2004high,liu2009virial} indicates a systematic error in this regime. This is possibly due to a breakdown of hydrodynamics during the time-of-flight as expected at high temperature.

\subsection{Fermi liquid behavior in the normal phase}
Above the superfluid transition and in the low-temperature regime $0.05<\zeta<0.5$, our data is well modelled by a Fermi liquid equation of state 
\begin{equation}\label{eq_Fermi}
P^{\mathrm{FL}}(\mu,T)=\frac{2}{15\pi^2}\left(\frac{2m}{\hbar^2}\right)^{3/2}\mu^{5/2}\left(\xi_n^{-3/2}+\frac{5\pi^2}{8}\xi_n^{-1/2}\frac{m^*}{m}\left(\frac{k_BT}{\mu}\right)^2\right),
\end{equation}
where $\xi_n=0.51(1)$ and $m^*=1.12(3)m$ respectively characterize the compressibility of the normal phase extrapolated to zero temperature and the effective mass of the low-lying excitations. The agreement with (\ref{eq_Fermi}) is better than $5\%$ in a large parameter range  $0.33\,\mu<k_BT<2\,\mu$. Our value of $\xi_n$ is in agreement with the
variational Fixed-Node Monte-Carlo calculations $\xi_n=0.54$ in \cite{carlson2003superfluid}, $\xi_n=0.56$ in
\cite{lobo2006nsp}, and with the Quantum Monte-Carlo calculation $\xi_n=0.52$ in \cite{bulgac2008quantum}. It is surprising that the quasi-particle mass $m^*$ is quite close to the free fermion mass, despite the strongly-interacting regime. Note also that this mass is close to the effective mass $m^*=1.20\,m$ of a single spin-down atom immersed into a Fermi sea of spin-up particles (the Fermi polaron) \cite{chevy2006upa,lobo2006nsp,combescot2007nsh,prokof'ev08fpb,combescot2008nsh,shin2008des,nascimbene2009pol,nascimbene2009eos}.

\subsection{Superfluid transition}
The deviation of the experimental data from (\ref{eq_Fermi}) for $\zeta<0.05$ signals the superfluid phase transition. This transition  belongs to the $U(1)$ universality class, and the critical region is expected to be wide \cite{taylor2009critical} in the unitary limit. Assuming that our low-temperature data belongs to the critical region, we fit our data with a function
\begin{equation}\label{eq_c}
P(\mu,T)=P^{\mathrm{FL}}(\mu,T)+A (\zeta_c-\zeta)^{2-\alpha} H(\zeta_c-\zeta),
\end{equation}
where $H$ is the Heaviside function and $\alpha\simeq-0.013$ is the specific heat critical exponent, measured with a very good accuracy on liquid $^4$He \cite{lipa1983very}. We obtain the position of the superfluid transition $\zeta_c=0.05$, or $k_BT_c/\mu=0.33$, in agreement with the value $k_BT_c/\mu=0.32(3)$ extracted in \cite{nascimbene2009eos} using a simpler fit function. We thus confirm more rigorously our previous determination of the superfluid transition. %, given the 10$\%$ systematic uncertainty. Associated 
In the appendix we discuss the validity of local density approximation around the superfluid transition. Under our current experimental conditions, the deviation from LDA is very small.

\section{Thermodynamics of a weakly-interacting Bose gas}
\begin{figure}[ht]
      \begin{center}
   \includegraphics[width=\linewidth]{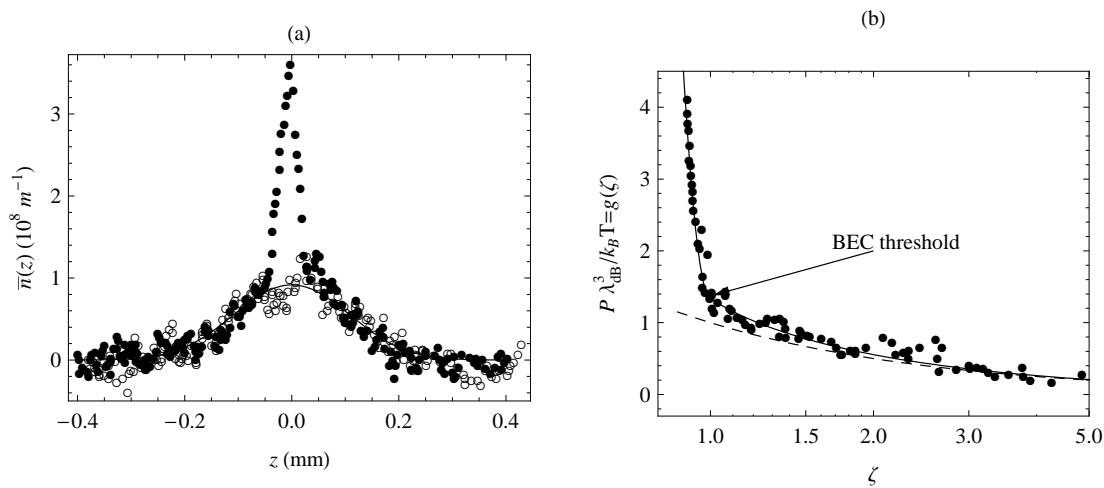}
   \vspace{-1.2cm}
      \end{center}
   \caption{(a) Integrated density profiles $\overline{n}(z)$ for the $^7$Li component (blacks dots) and the $^6$Li component (open circles). The solid line is a fit of the $^6$Li component with a finite-temperature Thomas-Fermi profile, yielding $T=1.6(1)~\mu$K. (b) Thermodynamic function $g(\zeta)$ determined from the $^7$Li profile. The solid line is a fit of the data with a Bose function in the non-condensed region and a mean-field equation of state in the condensed region (see text). The dashed line is the equation of state of a classical gas $g(\zeta)=\zeta^{-1}$. The difference between the dashed and solid lines around $\zeta=1$ is a consequence of Bose statistics.
\label{Fig_bec}}
\end{figure}

In this section we apply equation (\ref{Pressure}) to the case of trapped Bose gases. First we test the method by determining the equation of state of a weakly-interacting Bose gas \cite{caracanhas2009finite,romero2005equation}. We use an \emph{in situ} absorption image of a $^7$Li gas taken from \cite{schreck2001quasipure} (see Fig.\ref{Fig_bec}a). $^7$Li atoms are polarized in the internal state $|F=1,m_F=-1\rangle$, and held in an Ioffe-Pritchard magnetic trap with $\omega_r/2\pi=4970$~Hz and $\omega_z/2\pi=83$~Hz, in a bias field $B_0\simeq2$~G. Thermometry is provided by a gas of $^6$Li atoms, prepared in $|F=\frac{1}{2},m_F=-\frac{1}{2}\rangle$, and in thermal equilibrium with the $^7$Li cloud.

\subsection{Determination of the equation of state}
The equation of state of a weakly-interacting Bose gas can be expressed, in the grand-canonical ensemble, as:
\[
P(\mu,T)=\frac{k_BT}{\lambda_{dB}^3(T)}g(\zeta),
\]
where $\zeta=e^{-\mu/k_BT}$ is the inverse fugacity and $\lambda_{dB}(T)=\sqrt{2\pi\hbar^2/mk_BT}$ is the thermal de Broglie wavelength. The pressure profile is calculated using (\ref{Pressure}). We aim here at measuring $g(\zeta)$.
%The gas temperature $T=1.6(1)~\mu$K is determined from the density profile of a polarized $^6$Li gas at thermal equilibrium with the $^7$Li component (see Fig.\ref{Fig_bec}a). 
We obtain the global chemical potential value $\mu^0=0.10\,k_BT$ by fitting the $^7$Li profile in the non-condensed region $|z|>50~\mu$m with a Bose function:
\[
P(\mu_z,T)=\frac{k_BT}{\lambda_{dB}^3(T)}g_{5/2}(\zeta_z),\quad\zeta_z=e^{-\mu^0/k_BT}\exp\left(\frac{m\omega_z^2z^2}{2k_BT}\right),\quad g_{5/2}(z)=\sum_{k=1}^\infty \frac{z^{-k}}{k^{5/2}}.
\]
Combining the measurement of the pressure profile, of the cloud's temperature $T$ and global chemical potential $\mu^0$, we obtain the thermodynamic function $g(\zeta)$ plotted in Fig.\ref{Fig_bec}b. 

\subsection{Analysis of the equation of state}
In the region $\zeta>1$ the data agrees with the Bose function $g(\zeta)=g_{5/2}(\zeta)$ expected for a weakly-interacting Bose gas. The departure from the thermodynamic function of a classical gas $g(\zeta)=\zeta^{-1}$, and especially the fact that $g(\zeta)>1$ above the condensation threshold, is a thermodynamic signature of a bosonic bunching effect, as observed in \cite{yasuda1996observation,fölling2005spatial,schellekens2005hanbury}. The sudden and fast increase of our data for $\zeta\lesssim1$ indicates the Bose-Einstein condensation threshold. In the local density approximation framework, the chemical potential of a weakly-interacting Bose-Einstein condensate reads:
\[
\mu=\frac{4\pi\hbar^2a_{77}}{m_7}n,
\]
where $m_7$ is the $^7$Li atom mass and $a_{77}$ is the scattering length describing $s$-wave interactions between $^7$Li atoms. We neglect here thermal excitations in the condensed region. Integrating Gibbs-Duhem relation at fixed temperature $\mathrm{d}P=n\mathrm{d}\mu$ between the condensation threshold $\zeta_c$ and $\zeta<\zeta_c$, and imposing the continuity at $\zeta=\zeta_c$, we obtain the equation of state in the condensed phase:
\begin{equation}\label{g_BEC}
g(\zeta)= g_{5/2}(\zeta_c)+\frac{\lambda_{dB}(T)}{4\,a_{77}}(\log^2\zeta-\log^2\zeta_c).
\end{equation}
Fitting our data with the function $g(\zeta)$ given by (\ref{g_BEC}) for $\zeta<\zeta_c$ and with $g_{5/2}(\zeta)$ for $\zeta>\zeta_c$, we obtain $\zeta_c=1.0(1)$ and $a_{77}=8(4)a_0=0.4(2)$~nm. The uncertainties take into account the fit uncertainty and the uncertainty related to the temperature determination. The condensation threshold is in agreement with the value $\zeta_c=1$ expected for an ideal Bose gas, the mean-field correction being on the order of 1$\%$ \cite{giorgini1996condensate,giorgini1997thermodynamics}.
%The scattering length value is also in agreement with the calculation $a_{77}=11.5\,a_0$ in \cite{moerdijk1994prospects}.
Our measurement of the scattering length is in agreement with the most recent calculations $a_{77}=7(1)\,a_0$  \cite{servaas}.

Extending this type of measurement to larger interaction strengths on Bose gases prepared close to a Feshbach resonance would reveal more complex beyond-mean-field phenomena, provided thermal equilibrium is reached for strong enough interactions.

\section{Mott-insulator behavior of a Bose gas in a deep optical lattice}
Here we extend our grand-canonical analysis to the case of a $^{87}$Rb gas in an optical lattice in the Mott insulator regime. By comparing experimental data with advanced Monte Carlo techniques, it has been shown  that in many circumstances the local density approximation is satisfied in such a system \cite{trotzky2009suppression}. We analyze the integrated density profiles of the Munich group, Fig. 2 of \cite{fölling2006formation}.
%In this section we analyze the integrated density profiles of a $^{87}$Rb gas in an optical lattice from \cite{fölling2006formation}. 

\subsection{Realization of the Bose-Hubbard model with ultracold gases}
Atoms are held a trap consisting of the sum of a harmonic potential $V_h(x,y,z)$ and a periodic potential
\[
V_0\left(\sin^2(kx)+\sin^2(ky)+\sin^2(kz)\right),
\]
created by three orthogonal standing waves of red-detuned laser light at the wavelength $\lambda=2\pi/k=843$~nm.  The atoms occupy the lowest Bloch band and realize the Bose-Hubbard model \cite{jaksch1998cold}:
\begin{equation}\label{Hubbard_Hamiltonian}
\hat{H}=-J\sum_{\left<i,j\right>}\hat{a}_i^\dagger\hat{a}_j+\frac{U}{2}\sum_i(\hat{a}_i^\dagger\hat{a}_i-1)\hat{a}_i^\dagger\hat{a}_i,
\end{equation}
with a local chemical potential $\mu(\mathbf{r})=\mu^0-V_h(\mathbf{r})$. The index $i$ refers to a potential well at position $\mathbf{r}_i$, $J$ is the tunneling amplitude between nearest neighbors, and $U$ is the on-site interaction, $U$ and $J$ being a function of the lattice depth \cite{bloch2008many}. The slow variation of $V_h(\mathbf{r})$ compared with the lattice period $\lambda/2$ justifies the use of local density approximation.

We consider here the case of a large lattice depth $V_0=22E_r$, for which $J\simeq 0.003\, U\sim0$, and assume that the temperature is much smaller than $U$. In this regime the gas is expected to form a Mott insulator: in the interval $\mu\in[(p-1)U,pU]$, where $p$ is an integer, the atom number per site remains equal to $p$, and the density is equal to $n=p(2/\lambda)^3$. Integrating Gibbs-Duhem relation between 0 and $\mu$, we obtain that the pressure $P$ is a piecewise linear function of $\mu$:
\[
P(\mu,T=0)      = \left(\frac{2}{\lambda}\right)^3\left(\mu-\frac{p-1}{2}U\right)p\quad\mathrm{where}\quad(p-1)U<\mu<pU.\label{EOS_Mott}
\]

\subsection{Determination of the equation of state}
\begin{figure}[ht]
      \begin{center}
   \includegraphics[width=\linewidth]{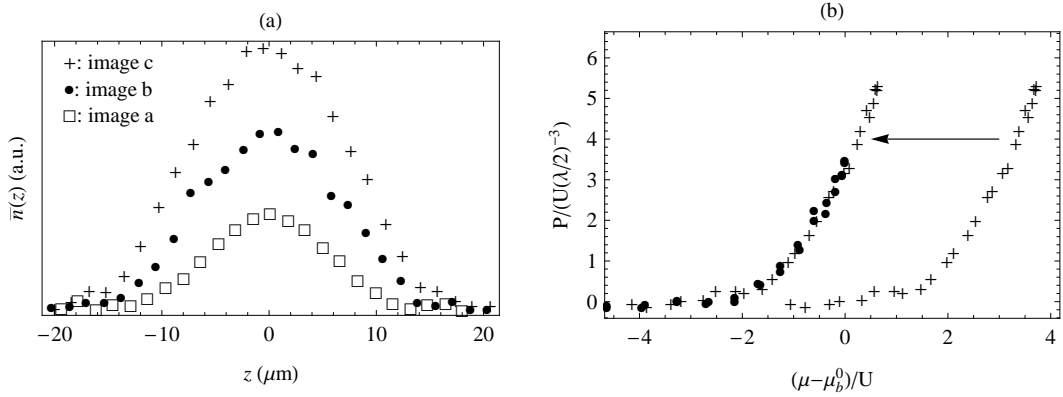}
   \vspace{-1cm}
      \end{center}
   \caption{(a) Integrated density profiles $\overline{n}(z)$ corresponding to image $a$ (open squares) $b$ (black dots) and $c$ (crosses), from \cite{fölling2006formation}. (b) Determination of the global chemical potential difference $\mu^0_c-\mu_b^0$ by superposing the equations of states given by each image. 
\label{Fig_Mott_construction}}
\end{figure}

We use a series of three images from \cite{fölling2006formation}, labeled $a$, $b$ and $c$, with different atom numbers  $N_a=1.0\times10^5$, $N_b=2.0\times10^5$ and $N_c=3.5\times10^5$ (see Fig.\ref{Fig_Mott_construction}a). The integrated profiles $\overline{n}(z)$ are not obtained using \emph{in situ} absorption imaging but rather using a tomographic technique,
%space-selective transfer of atoms at a given abscissa $z$ towards another internal state, which is imaged afterwards, 
providing a $\sim1~\mu$m resolution. The pressure profile is then obtained using equation (\ref{Pressure}).

\begin{figure}[ht!]
      \begin{center}
   \includegraphics[width=0.7\linewidth]{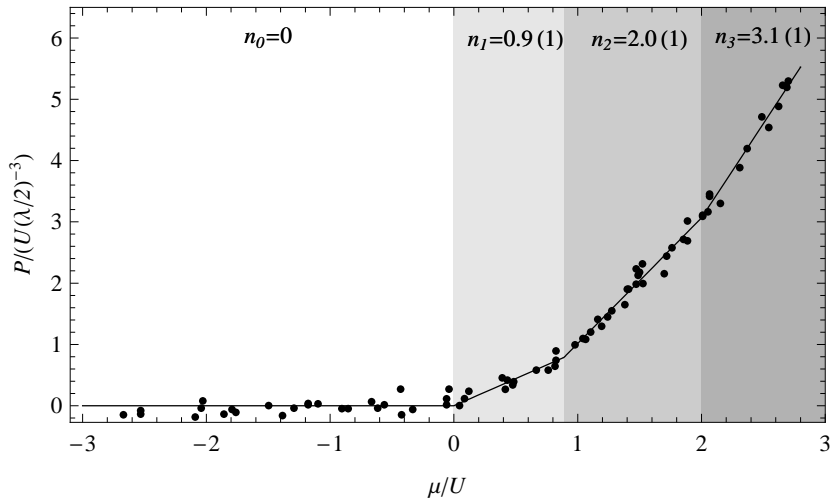}
   \vspace{-0.1cm}
      \end{center}
   \caption{Equation of state of a Bose gas in an optical lattice, in the Mott insulator regime. The solid line is a fit with a piecewise linear function capturing the Mott insulator behavior. The slope $\mathrm{d}P/\mathrm{d}\mu$ provides the density in each of the Mott zone, $n_1=0.9(1)$, $n_2=2.0(1)$, $n_3=3.1(1)$.
\label{Fig_Mott_EOS}}
\end{figure}

Each image $i=a,b,c$ plotted as $P$ as a function of $-\frac{1}{2}m\omega_z^2z^2$ provides the equation of state $P(\mu)$ translated by the unknown global chemical potential $\mu_i^0$. By imposing that all images correspond to the same equation of state (in the overlapping $\mu/U$ region), we deduce the chemical potential differences between the different images $\mu^0_b-\mu^0_a=0.56\,U$ and $\mu^0_c-\mu^0_b=0.61\,U$ (see Fig.\ref{Fig_Mott_construction}b). Gathering the data from all images, we thus obtain a single equation of state, translated by $\mu^0_a$ which is still unknown. 
We fit this data with a function translated by $\mu^0_a$ from the following function, capturing the Mott insulator physics:
\begin{eqnarray*}
\frac{P}{U(\lambda/2)^{-3}}&=&0\quad\mathrm{for}\quad \mu<0\\
                           &=&n_1\frac{\mu}{U}\quad\mathrm{for}\quad 0<\mu<\delta\mu_1\\
                           &=&n_1\frac{\delta\mu_1}{U}+n_2\frac{\mu-\delta\mu_1}{U}\quad\mathrm{for}\quad \delta\mu_1<\mu<\delta\mu_1+\delta\mu_2\\
                           &=&n_1\frac{\delta\mu_1}{U}+n_2\frac{\delta\mu_2}{U}+n_3\frac{\mu-\delta\mu_1-\delta\mu_2}{U}\quad\mathrm{for}\quad \delta\mu_1+\delta\mu_2<\mu,\\
\end{eqnarray*}
with $\mu^0_a$, $\delta\mu_1$, $\delta\mu_2$, $n_1$, $n_2$ and $n_3$ as free parameters. The value $\mu^0_a=1.51\,U$ yielded by the fit thus corresponds to the condition $P\rightarrow0$ when $\mu\rightarrow0$. Once it is determined, we obtain the equation of state of the Bose-Hubbard model in the Mott regime, plotted in Fig.\ref{Fig_Mott_EOS}. 

\subsection{Observation of a Mott-insulator behavior}
After fitting the value of $\mu^0_a$, the other parameters resulting from the fit exhibit the characteristic features of incompressible Mott phases. The occupation number in the first Mott region is $n_1=0.9(1)$ atom per site and the size is $\delta\mu_1=0.9(1)U$. The second Mott region occupation number is $n_2=2.0(1)$ and its size is $\delta\mu_2=1.1(1)U$. Finally, the third Mott region occupation number is $n_3=3.1(1)$. These values agree with the theoretical values $n_i=i$ and $\delta\mu_i=U$, in the $T=0$ and $J=0$ limits. 

\subsection{Estimation of finite temperature effects}
The equation of state deduced from the experimental data is also suited for investigating finite-temperature effects. Since sites are decoupled in the regime $J\ll U,k_BT$ considered in this study, the finite-temperature equation of state is easily calculated from the thermodynamics of a single site \cite{gerbier2007boson,capogrosso2007phase}: 
\begin{equation}\label{Mott_T_Fini}
P(\mu,T)=\frac{k_BT}{(\lambda/2)^3}\log\left(\sum_{p=0}^\infty\exp\left(-\frac{Up(p-1)/2-\mu p}{k_BT}\right)\right).
\end{equation}
Fitting now the experimental data with (\ref{Mott_T_Fini}) and $T$ and $\mu^0_a$ as free parameters, we deduce:
\[
k_BT=0.09_{-0.09}^{+0.04}\,U.
\]
This value is in agreement with a direct fit of the density profiles and number statistics measurements \cite{gerbier_private}.
This temperature is significantly smaller than the temperature $k_BT^*\simeq 0.2\,U$ at which the Mott insulator is expected to melt \cite{gerbier2007boson}. Second, this temperature should be considered as an upper limit because of its uncertainty on the low-temperature side. Indeed, the finite resolution of the images tends to smear out the sharp structure associated with Mott insulator boundaries, leading to an overestimation of the actual temperature. To overcome this limit, the spin-gradient thermometry proposed in \cite{weld2009spin} could be employed.

\section*{Summary and concluding remarks}
To summarize, we have shown on various examples of Fermi and Bose gas systems how  \emph{in situ} absorption images can provide the grand-canonical equation of state of the homogeneous gas. This equation of state is obtained up to a global shift in chemical potential and we have given several examples for its determination. The method relies on the local density approximation, which is satisfied in many situations, but notable exceptions exist such as the case of the ideal Bose gas. The equation of state given by this procedure allows direct comparison with many-body theories. While we have illustrated here this method on a single-component Bose gas and a spin-balanced Fermi gas, it can easily be generalized to multi-component gases. For instance the phase diagram and the superfluid equation of state of spin-imbalanced Fermi gases have been obtained in \cite{nascimbene2009eos,navon2010eos}. We expect this method to be very useful in the investigation of Bose-Bose, Bose-Fermi and Fermi-Fermi mixtures. Finally the equation of state of a Bose gas close to a Feshbach resonance may reveal thermodynamic signatures of beyond-mean-field behavior in Bose-Einstein condensates \cite{papp2008bragg}.

\section*{Acknowledgments}
We are grateful to Fabrice Gerbier and Kenneth Guenter for stimulating discussions.  We acknowledge support from ERC (Ferlodim), ESF (Euroquam Fermix), ANR FABIOLA, R\'egion Ile de France (IFRAF), and Institut Universitaire de France.

\section*{Appendix: validity of local density approximation}
Let us now discuss the validity of local density approximation around the superfluid transition in our experiment. Along the $z$ axis, the correlation length $\xi$ diverges around the transition point $z=z_c$ according to $\xi\sim k_F^{-1}|(z-z_c)/z_c|^{-\nu}$, where $\nu=0.67$ is the correlation length critical exponent, directly measured in \cite{donner2007critical}, and in agreement with $\nu=(2-\alpha)/3$. Local density approximation is expected to become inaccurate in the region $z_c-\delta z<z<z_c+\delta z$ where $\delta z$ is given by \cite{taylor2009critical,pollet2010criticality}:
\[
\delta z\sim\xi(z_c+\delta z),\quad\emph{i.e.}\quad\delta z\sim z_c(k_Fz_c)^{-1/(1+\nu)}.
\]
$z_c$ is on the order of the cloud size along $z$, and is much larger that $k_F^{-1}$ which is on the order of the inter-particle distance. Given the parameters of our experiments, $(k_Fz_c)^{-1/(1+\nu)}\sim 1\%$ and
%$(k_Fz_c)^{-1/(1+\nu)}\sim (k_FR_{\mathrm{TF}})^{-1/(1+\nu)}\simeq 1\%$ given the parameters of our experiment, 
the size $\delta z$ where local density approximation is invalid is very small. Given the noise of our data (a few $\%$), the deviation from local density approximation is thus negligible. Investigating the critical behavior at the superfluid transition, such as measuring the critical exponent $\alpha$, would be an interesting development for this method, as proposed in  \cite{pollet2010criticality}.

\section*{References}

\bibliographystyle{unsrt}
\bibliography{biblio}

\end{document}